\documentclass[preprint,12]{elsarticle}

\usepackage{epsfig}

\usepackage{amssymb}
\usepackage{amsthm}

\newcommand\TT{\rule{0pt}{3.2ex}}       % Top strut
\newcommand\BB{\rule[-1.8ex]{0pt}{0pt}} % Bottom strut
\newcommand{\blt}[1]{\textcolor{blue}{#1}} % Highlight in blue

\usepackage{soul}
\usepackage{lineno}

\usepackage{comment}
\usepackage{float}
\usepackage{tabularx}
\usepackage{adjustbox}
\usepackage{xcolor}
\usepackage{amsmath}
\usepackage{booktabs}

\renewcommand{\hl}[1]{\textcolor{black}{#1}} % Highlight text is in white background and black text

\journal{Mechanical Systems and Signal Processing}

\begin{document}

\begin{frontmatter}



%\title{Fault Diagnosis and Prognosis Capabilities for Wind Turbine Hydraulic Pitch Systems: An Overview}

\title{Fault Diagnosis and Prognosis Capabilities for Wind Turbine Hydraulic Pitch Systems}

\author{Alessio Dallabona\corref{cor1}\fnref{1}}
\ead{aldall@dtu.dk}
\cortext[cor1]{Corresponding author.}
\affiliation[1]{organization={Department of Electrical and Photonics Engineering, Technical University of Denmark},
            addressline={Elektrovej 326},
            city={Lyngby},
            postcode={2800},
            country={Denmark}}
\author[1]{Mogens Blanke}
\ead{mobl@dtu.dk}
\author[2]{Henrik C. Pedersen}
\ead{hcp@energy.aau.dk}
\author[1]{Dimitrios Papageorgiou}
\ead{dimpa@dtu.dk}
\affiliation[2]{organization={Department of Energy Technology, Aalborg University},
            addressline={Pontoppidanstræde 111},
            city={Aalborg},
            postcode={9220},
            country={Denmark}}

\begin{abstract}
Wind energy is the leading non-hydro renewable technology. Increasing reliability is a key factor in reducing the downtime of high-power wind turbines installed in remote off-shore places, where maintenance is costly and less reactive. Defects in the pitch system are responsible for up to $20\%$ of a wind turbine downtime. \hl{Thus, monitoring such defects is essential for avoiding it. This paper presents a generic assessment of the diagnosis capabilities in hydraulic pitch systems, which are used in high-power wind turbines. A mathematical model of the non-linear system dynamics is presented along with a description of the most frequent faults that occur. Structural analysis is used to assess which defects can be detected in the pitch system. The structural properties are furthermore explored to investigate the possibility of reducing the amount of sensors without compromising the fault diagnosis capabilities. Robustness to model uncertainty is finally addressed and generic principles for estimating the detectable magnitude of wear and tear are presented.}
\end{abstract}


\end{frontmatter}



\section{Introduction}\label{intro}

% Why is the topic of interest
Wind turbines production increased by a record of 273 TWh ($17\%$) in 2021\footnote{Data of 2022 is less meaningful because of the energy crisis, more details at \citep{ieaWorldEnergyOutlook2022}.}, the highest among all renewable power technologies. The same year the total production reads 1870 TWh, almost as much as all the other non-hydro renewable technologies combined. The Net Zero Emissions target by 2050 is reflected in the tendency to increase the turbine's size and power and to move towards off-shore systems \citep{ieaWindElectricityAnalysis2022}. Minimizing the downtime for such systems is a priority. In addition to the maintenance cost, one needs to consider the missing revenue associated with the energy not being produced during the repair. The offshore installation makes maintenance more challenging and, thus, longer in time. Studies like \citep{gayoFinalReportSummary2011,carrollFailureRateRepair2016,linigerReliableFluidPower2015} have shown that the main source of failure in wind turbines is associated with the pitch regulation system. The authors in \citep{chengInfluenceMechanicalFaults2023}, \citep{etemaddarResponseAnalysisComparison2016a} and \citep{leeFailureAnalysisHydraulic2020a} have shown how failures in such systems may have severe consequences. For high-power models, the pitch regulation system is electro-hydraulic. Compared to purely electrical pitch systems, electro-hydraulic systems offer higher power density, more robustness, and smoother performances \citep{walgernReliabilityElectricalHydraulic2023,palavicinoFaultsDiagnosticsElectrified2018, luReviewRecentAdvances2009}. The application of fault detection and accommodation schemes can be leveraged to reduce the required maintenance on the hydraulic subsystem such that the up-time of the wind turbine significantly increases.

% What was done up to now
A significant corpus of literature has been focusing on the design and evaluation of methods for the detection and mitigation of individual faults that can occur in wind turbines. A comprehensive review of the research methods is presented in \citep{badihiComprehensiveReviewSignalBased2022, fekihFaultDiagnosisFault2022,fengReviewVibrationbasedGear2023,wangDVGTformerDualviewGraph2024,yucesanHybridPhysicsinformedNeural2022}. Actuator faults have also been addressed in the context of fault-tolerant control in \citep{songFaultTolerantControlFloating2023,palanimuthuReliabilityImprovementLargescale2023,liuActuatorFaultTolerant2023,mousaviFaulttolerantOptimalPitch2022}. The authors in \citep{asmussenFaultDetectionDiagnosis2021} focused on pitch system defects. It reviewed results from pitch system fault detection and also mentioned possibilities with fault-tolerant control. Other investigations related to fault diagnosis and mitigation for the pitch system included \citep{elorzaSensorDataProcessing2022a,venkaiahPitchControlElectrohydraulic2023,lanReviewFaultDiagnosis2023a} and the references therein. The hydraulic circuit driving the pitch system in wind turbines is similar to the ones used in several other applications for heavy machinery, such as mechanical presses and servo control of ship's rudders and variable pitch propellers. Several studies applied tools to detect and mitigate mainly leakage faults. The most recent and relevant results for detection/estimation and mitigation appear in \citep{goharriziWaveletBasedApproachExternal2011,vasquezActiveFaultDiagnosis2019,aslAdaptiveSquarerootUnscented2019,bahramiAdaptiveSupertwistingObserver2018a,shanbhagFailureMonitoringPredictive2021,choFaultDetectionDiagnosis2021,zhaoFaultDiagnosisHydraulic2020,liuFaultDiagnosisHydraulic2014,goharriziInternalLeakageDetection2012,xuModelBasedFaultDetection2015,sharifiMulticlassFaultDetection2016,sunMultiFaultDiagnosisApproach2020,zhangSARPerformancebasedFault2022,djordjevicSensorFaultEstimation2022,daiSignalBasedIntelligentHydraulic2019,shenFaultDiagnosisAircraft2022,ghanbariDetectionFaultsElectroHydrostatic2022,tongRootCauseDetection2023} and \citep{maddahiPracticalApproachDesigning2020,daoActiveFaultTolerant2021,phanOptimizedBasedFaultTolerantControl2022a,phanActuatorFailureCompensationbased2023,phanFaulttolerantControlElectrohydraulic2023,zhangMultiModelBasedAdaptive2022}.  The above references consider results for model-based, statistical-based, and learning-based methods, with focus on specific selected defects.

% What was attempted
Employing graph-based methods of topology, this paper provides a holistic investigation that systematically assess capability detecting and isolate defects in a hydraulic pitch system, and also addressing what is the coverage of diagnosis with the typical sensor configuration in a high-power wind turbine. The contributions of the paper are:
\begin{itemize}
    \item Include faults essential for resilience and maintenance in a generic benchmark model.
    \item Employ structural analysis of the model to assess generic capabilities for diagnosis. 
    \item Investigate sensor topology aiming at cost reduction while retaining coverage of diagnosis.
    \item \hl{Assess the robustness of fault detection to model uncertainty.}
\end{itemize}

% What is presented in the paper
The remainder of the paper is structured as follows: Section~\ref{ch2} defines a benchmark model that is used to investigatecondition monitoring and fault diagnosis in hydraulic pitch systems. The fault detection and isolation capabilities of the system are analyzed in Section~\ref{ch3} by carrying out a structural analysis on the benchmark model. Section~\ref{ch4} investigates topology alterations that could improve fault diagnosis in the system. \hl{Robustness of fault detection with respect to model uncertainty is treated in \mbox{Section~\ref{new5}}.} The results are discussed in Section~\ref{ch5}, and conclusions are drawn in Section~\ref{ch6}.

%%%%

\section{Pitch System Topology and Modeling}\label{ch2}

\begin{figure}[t]
\includegraphics[width=7.9 cm]{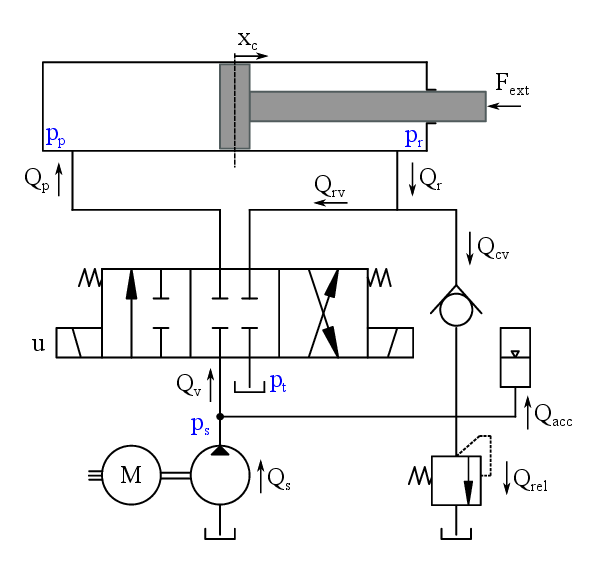}
\caption{Hydraulic pitch system components \citep{pedersenInvestigationLoadReduction2016}.\label{sys_mod}}
\end{figure}

The reference hydraulic pitch system for the article is presented in Figure~\ref{sys_mod} and represents a simplified schematics/topology of a pitch system under normal operation, but excluding the safety features. A pump and an accumulator bank (represented as one accumulator) are used to supply three equivalent circuits composed of a proportional valve (PV) and a hydraulic cylinder. More information about how such a system works is provided in \citep{linigerDesignReliableFluid2018}.

The benchmark model comprises the nominal dynamics detailed in \citep{pedersenInvestigationLoadReduction2016} and all major faults that could affect turbine availability and production. Although modeling for fault-tolerance may require a multiplicative representation of some defects \citep{blankeDiagnosisFaultTolerantControl2016}, all faults in this study are expressed as additive, time-varying terms (in blue color) since the effect of any generic fault can be described in this way when considering diagnosis. A thorough description and models for diagnosis are presented in Section~\ref{ch2_6}, and sections ~\ref{ch2_1} to ~\ref{ch2_3} present the detailed mathematical models for the elements of the pitch system.

\subsection{Hydraulic Cylinder}\label{ch2_1}

The main component allowing the pitching of the blades is the hydraulic cylinder. Its dynamics is described by the force balance:
\begin{linenomath}
\begin{equation}\label{eq_xc}
    M_{eq}\dot{v_c} = A_pp_p-A_rp_r-  B_v v_c - F_c\,tanh(v_c/\gamma)    \,
\textcolor{blue}{+ f_{Fr,c}}-F_{ext}
\end{equation}
\end{linenomath}

where the notation generally refers to Figure~\ref{sys_mod} and $f_{Fr,c}$ is an incipient fault due to increased friction. 

Dynamical relations likewise give the pressure build-up in the piston-side and the rod-side chambers: 
\begin{equation}\label{eq_pp}
    \dot{p_p} = \frac{\beta_{e,p}}{V_{0,p}+A_px_c}(Q_p-A_pv_c -Q_{le,p}-Q_{li})
\end{equation}
\begin{equation}\label{eq_pr}
    \dot{p_r} = \frac{\beta_{e,r}}{V_{0,r}+A_r(x_{c,max}-x_c)}(-Q_r+A_rv_c-Q_{le,r}+Q_{li}),
\end{equation}

where $\beta_{e,i}$ is the effective bulk modulus of the oil for the respective chamber. The bulk modulus of a fluid describes its ability to resist compression. Such property is affected by the presence of air in the fluid. The effective bulk modulus is introduced to account for this effect and it is described by the following pressure-dependent equation:
\begin{equation}\label{bmeq1}
    \beta_{e} = \frac{1}{\frac{1}{\beta_{oil}} + \varepsilon_a(p)\left(\frac{1}{c_{ad}p} - \frac{1}{\beta_{oil}} \right)  }\textcolor{blue}{+ f_{B,e}}
\end{equation}
\begin{equation}\label{bmeq2}
    \varepsilon_a(p) = \frac{1}{\frac{1-\varepsilon_{a,0}}{\varepsilon_{a,0}}\left(\frac{p_{atm}}{p}\right)^{\frac{-1}{c_{ad}}}+1} ,
\end{equation}
where the fault $f_{B,e}$ is related to oil degradation.

\hl{An example of the effective bulk modulus dependency on air content, pressure, and temperature is described in \mbox{\citep{knezevicAnalysisChangesBulk2011}}. Of the three, the air content is by far the most significant, but as the system normally will be operating with pressures well above 25 bar, changes in the effective bulk modulus will be very limited, even for air content up to 10\% volumetric air content (measured at atmospheric pressure). Therefore, the model does not consider temperature as a state variable. }

\hl{The high degree of uncertainty in the bulk modulus model described in \mbox{Equations~\ref{bmeq1} and \ref{bmeq2}}, and the fact that the pressures in the cylinders' chambers as well as in the supply circuit vary within similar values, has motivated the adoption of a single uncertain parameter $\beta_e$ to describe the effective bulk modulus in the entire circuit. This simplification is consistent with the fact that the fault $f_{B,e}$ is related to the oil degradation, regardless of the part of the plant where it appears. Thus, the simplified effective bulk modulus description is givne by the following equation:}

\begin{equation}
    \beta_{e} = \beta_{e,0}\textcolor{blue}{+ f_{B,e}}
\end{equation}

The remaining terms in Equations~\ref{eq_pp} and~\ref{eq_pr}  are the flow rates entering and exiting from the chambers ($Q_p$ and $Q_r$), and the leakage flows. The former are determined by the valve position and will be discussed later. The latter include an external leakage from each chamber and an internal one between the two of them:
\begin{equation}
    Q_{le,p}  = C_{le,p}(p_p -p_{atm}) \textcolor{blue}{+ f_{Q_{le,p}}}
\end{equation}
\begin{equation}
    Q_{le,r}  = C_{le,r}(p_r -p_{atm}) \textcolor{blue}{+ f_{Q_{le,r}}}
\end{equation}
\begin{equation}
    Q_{li}  = C_{li}(p_p -p_r) \textcolor{blue}{+ f_{Q_{li}}}.
\end{equation}

Although cylinder leakage flows are negligible in the nominal plant, there may be a minor cross-port leakage in the proportional valve (equivalent to internal leakage in the cylinder). Therefore, modeling leakage in the above three cases can be useful for separating the related faults $f_{Q_{le,p}}$, $f_{Q_{le,r}}$ and $f_{Q_{li}}$ in three distinct equations.

\subsection{Valves}\label{ch2_2}

The position of the cylinder is controlled by using a proportional valve (PV). By varying the valve's spool position, it is possible to regulate the flow entering in the hydraulic cylinder chambers, performing the pitching action. The equations describing the flows with respect to the valve's spool position are:
% \begin{equation}\label{eq_qp}
%     Q_p = K_v(x_v) \left[ \sqrt{|p_s-p_p|}sign(p_s-p_p)H(x_v) -  \sqrt{|p_p-p_t|}sign(p_p-p_t)H(-x_v)  \right]\textcolor{blue}{+ f_{Q_{p}}}
% \end{equation}
\begin{align}\label{eq_qp}
    Q_p = K_v(x_v) &\left[ \sqrt{|p_s-p_p|}sign(p_s-p_p)H(x_v) \right. \\
    &\left. - \sqrt{|p_p-p_t|}sign(p_p-p_t)H(-x_v)  \right]\textcolor{blue}{+ f_{Q_{p}}}
\end{align}
\begin{equation}\label{eq_qrv}
    Q_{rv}  = - K_v(x_v)\phi_v  \sqrt{|p_s-p_r|}sign(p_s-p_r)H(-x_v)\textcolor{blue}{+ f_{Q_{rv}}},
\end{equation} where $K(x_v)$ is the tabular function with which the manufacturer describes the input-to-flow relation of the controllable valve.The Heaviside step function $H(\cdot)$ distinguishes the different operating regions of the valve, i.e.,
\begin{equation}
    H(x) = \begin{cases}
    1, & x\geq 0\\
    0, & x< 0.
\end{cases}
\end{equation}

Generally, the flow can be affected by a malfunctioning valve. Hence, the possible faults $f_{Q_{p}}$ and $f_{Q_{rv}}$.

The valve is usually supplied by the manufacturer with an inner control loop that sets the valve's spool position to follow a reference, which is the result of the upstream controller. The resulting valve closed-loop dynamics is
\begin{equation}\label{eq_xv}
    \dot{v_v}+2\xi \omega_0\,{v_v}+\omega_0 ^2\,x_v \textcolor{blue}{+ f_{Fr,v}}= \omega_0^2\,x_{v,ref}\textcolor{blue}{+ f_{wv,v}} , \quad x_{v,ref}=k_uu .
\end{equation}

% \begin{equation}
%     \dot{x}_v = v_v
% \end{equation}

Two possible faults can affect the valve: a mechanical fault similar to the cylinder case, and an electrical one affecting the coils responsible for moving the valve's spool. The faults are kept separate to remark their different nature.

Equation~\ref{eq_qrv} describes the flow across the PV, which is generally different than the flow to/from the rod-side chamber, given by
\begin{equation}
    Q_{r}  = Q_{rv} + Q_{cv}.
\end{equation}

$Q_{cv}$ describes the flow through the check valve used for operating the circuit in regenerative mode, for reducing the size of the supply circuit. The equation for the flow in the check valve is given by
\begin{equation}\label{eq_qcv}
    Q_{cv}  = K_{cv} (p_r-p_s-p_{cv,c})H(p_r-p_s-p_{cv,c})\textcolor{blue}{+ f_{Q_{cv}}}.
\end{equation}

The valve can degrade or break. Thus, it introduces a possible fault $f_{Q_{cv}}$ manifesting as flow perturbation.

\subsection{Supply Circuit}\label{ch2_3}

Each cylinder-valve system is supplied by a common pressure source which is located in the nacelle of the turbine. From a representation point of view, the system is composed of a fixed displacement pump, an accumulator, and a pressure relief valve ensuring that the system's pressure is kept below a certain limit. The pump charges the accumulator until the nominal system pressure is reached, and its flow is then circulated through a filter back to the tank. Hence, the pressure source during nominal operation is the accumulator. As a consequence, the system's pressure varies over time. The equation describing the pressure dynamics at the accumulator is 
\begin{equation}\label{eq_ps}
    \dot{p_s} = \frac{1}{\frac{V_{oil}}{\beta_{eff,s}} + \frac{V_{gas}}{k\,p_s} }Q_{acc},
\end{equation}

where $Q_{acc}$ is the flow the accumulator exchanges with the rest of the system, given by
\begin{equation}\label{eq_qacc}
    Q_{acc} = Q_s - Q_{rel} + \sum_{i = 1}^{3}Q_{cv,i} - \sum_{i = 1}^{3}Q_{v,i}\,\textcolor{blue}{+ f_{Q_{ru}}}.
\end{equation} The flow $ Q_{v}$ from the supply circuit to each of the proportional valves is
\begin{equation}
    Q_{v} = Q_pH(x_v) - Q_{rv}H(-x_v).
\end{equation}

The flow $Q_{rel}$ through the relief valve, when the nominal pressure of the plant is overtaken, is given by
\begin{equation}\label{eq_qrel}
    Q_{rel}  = K_{rel} (p_s-p_{cr,r})H(p_s-p_{cr,r}) \textcolor{blue}{+ f_{Q_{rel}}}.
\end{equation}

Similarly to the other valves, the fault $f_{Q_{rel}}$ can be associated with its rupture.

Finally, the volumes of oil and gas in the accumulator are given by
\begin{equation}
    V_{oil} = V_{acc} + V_{hose} - V_{gas}
\end{equation}
\begin{equation}\label{eq_vgas}
    V_{gas} = V_{acc}\left[\left(\frac{p_{gas,0}}{p_s}\right)^{1/k}H(p_s-p_{gas,0})\right] \textcolor{blue}{+ f_{acc}}.
\end{equation}

A faulty accumulator can lead to gas leaking into the oil. For a given pressure, a reduction in the gas quantity is hence equivalent to a reduction in the volume it occupies. Thus, the effect of this fault is modeled as a volume perturbation $f_{acc}$.

\subsection{Sensors}

Generally, pressures in the chambers of each piston can be measured along with the position of the rod.

Different sensors can be used for monitoring the plant. Generally, for each piston, the position of the rod and pressures in the two chambers can be measured. Each PV is provided with a closed-loop control circuit, which leverages a position measurement for the control action to be established. Potentially, such signals can be extracted and utilized for fault diagnosis purposes. Finally, the pressure of the system can be measured. Other sensors, such as e.g. flow rate sensors, exist and are possible to be installed. However, they are not economically feasible. Hence, they are excluded from the analysis.

The addition of each sensor implies the possibility that each of them can fail, meaning that an equal number of faults need to be included in the model.

\subsection{Fault Description}\label{ch2_6}

In the previous subsections, faults have been inserted as additive terms in the equations describing the nominal plant operation. For setting up a structural analysis, it is important to specify where the fault enters, as that is the only relevant information for evaluating whether it is possible to detect it or isolate it from other faults. In reality, since different faults have different sources, they also have different properties as signals. This information is only relevant in the design phase of the diagnostic tools. Table~\ref{table_faults} presents a basic description of all the faults according to \citep{papageorgiouOnlineFrictionParameter2020,asmussenFaultDetectionDiagnosis2021,choNumericalModelingHydraulic2020,linigerReliabilityBasedDesign2017,shanbhagFailureMonitoringPredictive2021,khanFaultDetectionElectrohydraulic2002}.

\begin{figure}[h]
\includegraphics[width=12.5 cm]{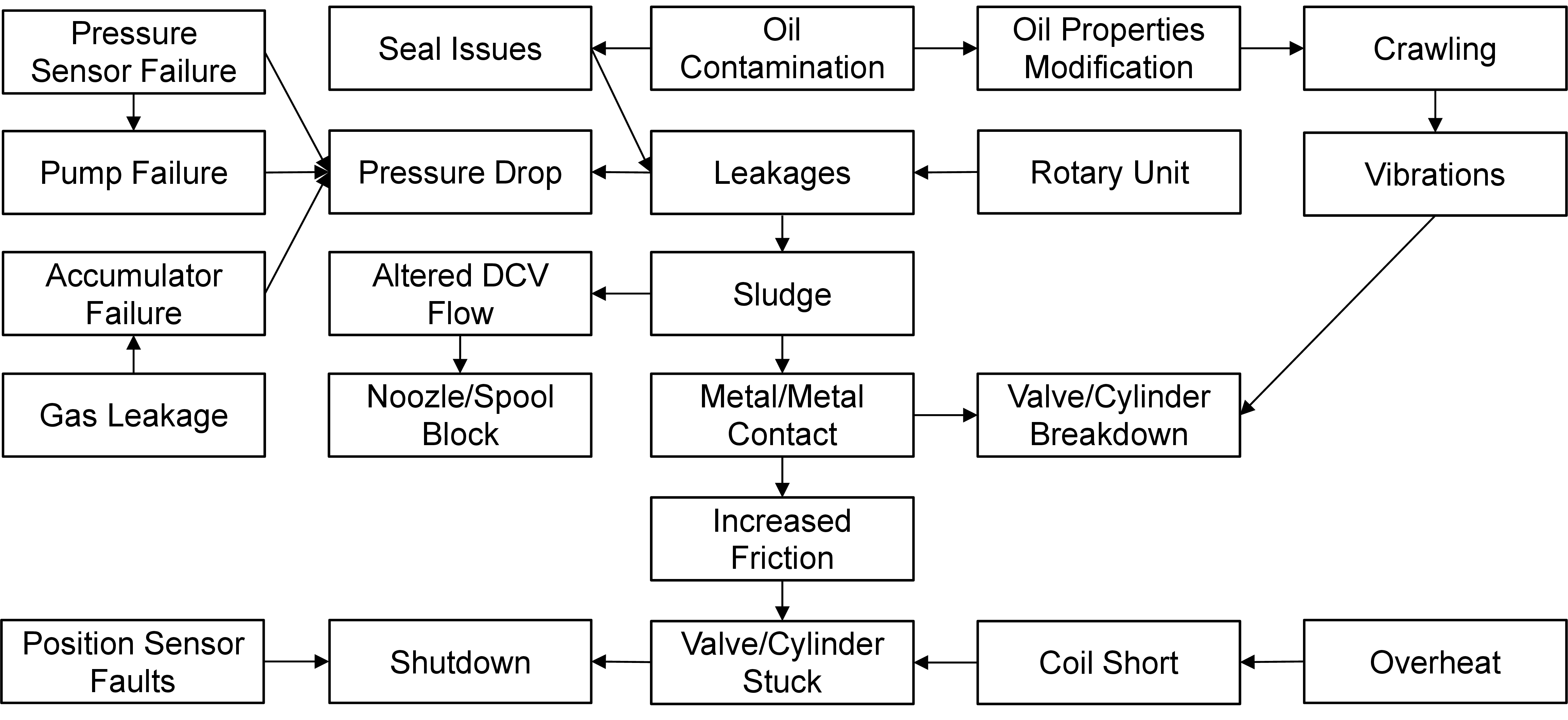}
\caption{Diagram of most relevant fault modes and their causes.}\label{fault_tree}
\end{figure}

The description column provides information on how different faults are related. The severity of a given fault depends on how it propagates in the system. Figure~\ref{fault_tree} illustrates such propagation properties of the faults listed in the table.

The mathematical model and Table~\ref{table_faults} do not include description related to pump failure, as there is no dynamical relation associated with it. A fault in the pump will, from an operational point of view, typically result in reduced pump flow and ripples, leading to slower pressure build-up in the system. Detection of such a fault is obtained by pump pressure monitoring using a pressure sensor in the vicinity of the pump.

\begin{table}[H]
\newcolumntype{C}{>{\centering\arraybackslash}c}
\caption{Failure Modes. The rating of failures when a component specific  threshold is exceeded. \label{table_faults}}
    \centering
    \begin{adjustbox}{max width=1\textwidth}
    \begin{tabular}{CCm{7.5cm}C}
         \toprule
			\textbf{Fault Name}	& \textbf{Fault Model}  & \textbf{Description} &  \textbf{Severity}\\
			\midrule
    $f_{Fr,c}$	& $\begin{aligned}
        &\Delta_{B_c}v_c\\
        &+ \Delta_{F_c}\,tanh(v_c/\gamma)
    \end{aligned}$	& Increased friction in the cylinder because of sludge formation. Not critical under a certain threshold, but it can cause the cylinder to get stuck. In the initial phase it appears as a variation of friction coefficients. & High\\
        \midrule
    $f_{B,e}$ & $\Delta_{B,e}$ 					& Fault related to oil (i.e. gas) contamination, difficult to model differently than a generic variation with respect to the nominal value. & Medium\\
        \midrule
    $f_{Q_{le,p}}\, f_{Q_{le,r}} $		&		 $\Delta_{C_{le,i}}(p_i -p_{atm})$			& Oil leakage between a cylinder chamber and the outside. It can be slow or abrupt, according to wear. Even with small values the turbine will be shut down. From a control point of view, leakage do however act to increase the damping in the system. & High\\
    \midrule
    $f_{Q_{li}}$	& $\Delta_{C_{li}}(p_p -p_{r})$					& Oil leakage between the two chambers in the cylinder. Same characteristics as the external one, it just increases damping in the pitching. & Low\\
    \midrule
    $f_{Q_{p}}\, f_{Q_{rv}}$	& $\Delta_{K_v}[\cdot]$	& Variation in the proportionality coefficient due to mechanical wear of the valve, or by incorrect valve command. As long as the valve is able to operate the severity is low.  & Low\\
    \midrule
    $f_{Fr,v}$	&   $\begin{aligned}
        &2\Delta_\xi \Delta_{\omega_0}\,{v_v}\\
        &+\Delta_{\omega_0} ^2\,(x_v-x_{v,ref})
    \end{aligned}$			& Emergence of significant friction in the valve because of sludge formation or wear. The effect is seen in the closed-loop dynamics parameters, and may lead to the valve getting stuck with time. & High\\
    \midrule
    $f_{wv,v}$	& $\omega_0^2\Delta_{k_u}u$ & Fault in the coil, leading to a wrong voltage setting. It is modeled as a wrong position setting since the impact is equivalent. & High\\
    \midrule
    $f_{Q_{cv}}$	& $\Delta_{Q_{cv}}$ & It's usually a discrete-time nature fault leading to the valve getting stuck, more than leakages or a variation in the proportional flow coefficient. & High\\
    \midrule
    $ f_{Q_{ru}}$	& $ \Delta_{Q_{ru}}$ & Leakage flow in the rotary unit, i.e., the connection between the pressure source and the three actuation circuits. & Low\\
    \midrule
    $f_{Q_{rel}}$	& $\Delta_{Q_{rel}}$				& It can be a variation in the proportional coefficient, leakage, or a variation in the pressure at which it is triggered, although these type of faults are rare. & Low\\
    \midrule
    $f_{acc}$	& $-V_{gas,l}$	& Gas leakage in the accumulator, with consequent reduction of the volume of gas, for a given pressure. Generally slowly varying. & Medium\\
    \midrule
$f_{y,p_s}$\,$f_{y,p_p}$\,$f_{y,p_r}$	& $\Delta_{y,p_i}$					& Wrong measurement in the related pressure sensor. & Low\\
    \midrule
    $f_{y,x_c}$\,$f_{y,x_v}$& $\Delta_{y,x_i}$						& Wrong measurement in the cylinder/valve position sensor. Position measurements are the only signals used in the control loop. Hence, their importance. & High\\
               \bottomrule
    \end{tabular}
    \end{adjustbox}
    \label{tab:my_label}
\end{table}

\vspace*{0.3cm}

%%%%

\section{Structural Analysis of the Benchmark Model}\label{ch3}

This section provides an introduction to the key structural properties of dynamical systems that are used for fault detection and presents a systematic assessment of the system's diagnostic capabilities. The structural analysis is first performed at a generic operating point and later separately applied to all the individual operating regions.

\subsection{\hl{Analysis based on system's structure}}

\hl{The modelling needed for fault-tolerant design pertains to the topology of the system and is based on the individual relations between the variables. This is done through formulation of \emph{constraints}. When modelling to obtain overall functional relations, it is desirable not to predefine what is input and what is output of individual function blocks, but rather state that there are relations between these variables, constrained by given relations. Such constraints were described in the modelling Sections \mbox{\ref{ch2_1}} to \mbox{\ref{ch2_3}}.}
\hl{Modelling for fault-tolerance is therefore conveniently done using the principles of behavioural modelling, where constraints $c$ describe how variables are related. Let variables be $x, z$ where $x$ are unknown, $z$ are known (input $u$ or output $y$), and disturbances are $d$. Let $g_{s}$ and $g_{d}$ denote functions; then constrains can be static $(c_s)$ or dynamic $(c_d)$ :}
\begin{equation*}
\begin{array}{c c c c}
  c_{s}: &0 & = & g_{s}(x,d) - z \\
  c_{d}: &0 & = & g_{d}(x,d) - \dot{x},\\
\end{array}
\end{equation*}
\hl{where $g_s$ and $g_d$ can be linear or nonlinear. Derivatives of variables can be explicit or implicit in
the constraints. }

A violation of one or more constraints is an indication that some fault has occurred, as the model is not able to describe the correct dynamics anymore. By combining the information of the constraints with the knowledge of the possible faults that may enter the plant, it is possible to understand which of the latter are detectable and, eventually, isolable, for a given plant configuration (e.g. sensor placement list). It is worth noting that these properties only depend on the system's structure and not on the design of the diagnostic system. Thus, a structural analysis advises on possibilities for detectability and isolability of faults in a system based on the knowledge of its structure. This provides us with its generic \emph{structural detectability and isolability} properties. 

Structural analysis is a method where a graph represents system topology. The basis is a graph consisting of: 1) function blocks, each of which represents constraints that describe behaviors of the function block; 2) variables in the system; 3) paths that describe links between constraints and variables. A system's topology is described as a graph having constraints as the set of vertices a), variables in the system, b), and edges c) that connect the two types of vertices. The essence of structural analysis is to use tools from the arsenal of graph theory to obtain causal matchings in the bi-partitioned graph. A matching consists of a set of paths between unknown variables and constraints \citep{dulmageCoveringsBipartiteGraphs1958}. When an unknown variable $x$ (e.g. a state or a disturbance) is matched to a constraint $c$, the latter can be used to calculate the unknown variable. When all unknown variables are matched to constraints, any additional unmatched constraint can be used to describe redundant information to compute so called residuals. Residuals would be zero when there are no violations of normal behaviors and some will be non-zero if one or more behaviors are violated. Structural models were described at a tutorial level in \citep{blankeStructuralDesignSystems2007} and references herein and software tools are available for swift analysis of structural properties. The SaTool \citep{blankeSaToolSoftwareTool2006} focuses on violation of constraints and residual generation, including the computationally efficient method in \citep{krysanderEfficientAlgorithmFinding2008a}.

Based on its graph-based representation, a system can be decomposed into three subsystems, i.e., three sub-graphs. The \emph{over-determined} part contains unknown variables that are all matched as well as unmatched constraints. The \emph{just-determined} subsystem contains equal number of constraints and unknown variables, which are matched to each other. Finally, the \emph{under-determined} part contains unmatched unknown variables. A necessary condition for fault detection is that the over-constraint subsystem is nonempty and the under-constraint subsystem is the empty set. The Dulmage-Mendelson (DM) algorithm \citep{dulmageStructureTheoryBipartite1959} is the original method for obtaining the structural decomposition of a system. Constraints are divided into just-determined and over-determined parts. If a fault appears in the just-determined part, no redundancy is present, and it is not detectable. The redundant equations in the over-determined part can be used for detecting faults that are affecting them. As a result, faults in the over-determined part are structurally detectable. Determining whether a fault is also isolable, requires a further analysis that includes looking into generation of \emph{residuals}. An efficient way of determining the isolability properties was developed in \citep{chenEfficientMethodDetermining2021}.

%%%%%%%%%%%%%%%%%%%%%%%%%%%%%%%%%%%%%%%%%

% Connecting sentence

\hl{Analytical redundancy relations are found by using the minimal structural over-determined subsets (MSO) and by backtracking through a selected matching, $\mathcal{M}$. Starting with $c_{n+1}$ being an unmatched constraint, whereas $c_{n}$ to $c_{1}$ are matched, $c_1,...,c_n \in \mathcal{M}_{N}$, then, with $z_i \in \mathcal{Z}$ where $\mathcal{Z} = \{x_i,d_j,f_k\}$ is the set of all unknown variables $x_i$, disturbances $d_j$, explicitly modelled faults, $f_k$, known input $u_n$, and measurements $y_m$,}
\begin{align}\label{ARR}
c_{N+1}(z_1,..z_N) = c_{N+1}(c_1(\{x_i,d_j,f_k,u_n,y_m\} \in \mathcal{Z}_1, \\ \nonumber  ...,c_N(\{x_i,d_j,f_k,u_n,y_m\}\in \mathcal{Z}_N),
\end{align}
\hl{where the set \mbox{$\mathcal{Z}_1 \subset \mathcal{Z}$} is the parts of the sub-graph that are employed to match \mbox{$c_1$}, etc.}

\hl{A residual $r$ is obtained from $c_{N+1}$ by iteration through the matching $\mathcal{M}_1$ that calculates the variables $x,d,f$ such that the residual becomes a function solely of the known variables $u,y \in \mathcal{M}_1$, i.e., a residual is,}
\begin{equation}\label{residual 1 ideal}
    r(t) = g(\{c(\cdot), u(t), y(t)\} \in \mathcal{M}_1),
\end{equation}
\hl{which is evaluated in real time.}

\hl{This construction of residuals from structural analysis guarantees that violation of any constraints are structurally detectable. It does not express the sensitivity with which a particular residual can reflect a violation of constraints, i.e. fault(s) in the system since sensitivity is not a structural property, but a property in the analytic domain.}

\hl{Design of diagnosis and monitoring algorithms therefore has two steps. One is analysis of structure that shows which possibilities of change detection exist with a given system topology. The second step is to  materialize this in the analytic domain where sensitivity to changes (faults) and robustness to disturbances and parameter uncertainty are addressed.}

%%%%%%%%%%%%%%%%%%%%%%%%%%%%%%%%%%%%%%%%%%%%%%%%%%%%%%%%%%%%%%%%%%%%

\subsection{Full Model Analysis}\label{ch3_1}

\begin{figure}[h]
\centering
\includegraphics[width=0.9\textwidth]{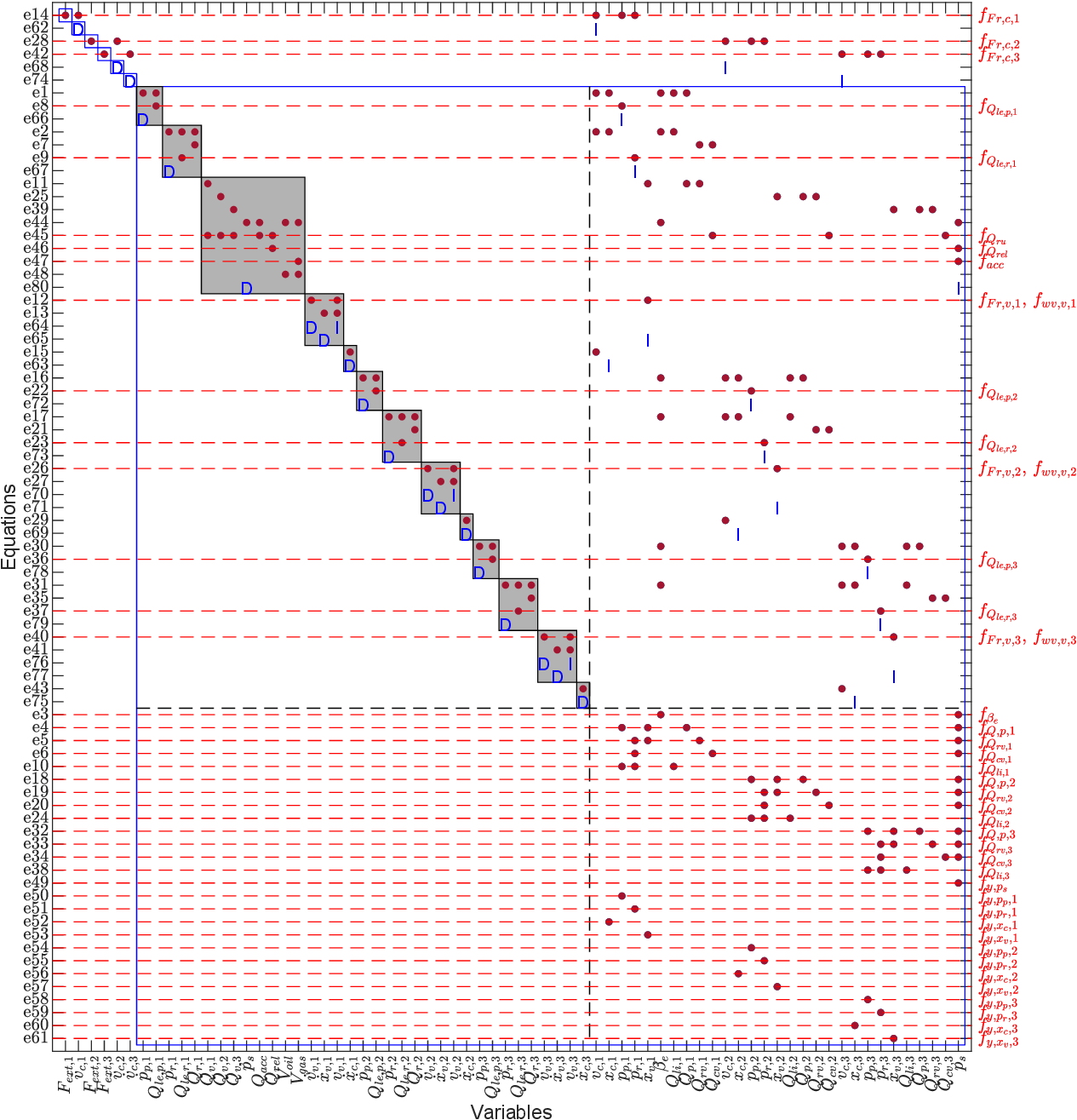}
\caption{DM Decomposition for the full plant.\label{dm_full}}
\end{figure}

The analysis is carried out by referring to the equations stated in Section~\ref{ch2}, with the support of the 'Fault Diagnosis Toolbox' from \citep{friskToolboxAnalysisDesign2017}. The standard sensors present in a real system are utilized. The analysis hence delivers a benchmark result for the capabilities of diagnosis in such a plant. Different sensor configurations will be tested in Section~\ref{ch4} to obtain an equipment topology that maintains the diagnostic features of the original one, but with a reduced set of sensors. The Dulmage-Mendelson's Decomposition for the standardized hydraulic pitch system, together with the partition in equivalent classes, is presented in Figure~\ref{dm_full}. 

\begin{figure}[tp]
\centering
\includegraphics[width=0.9\textwidth]{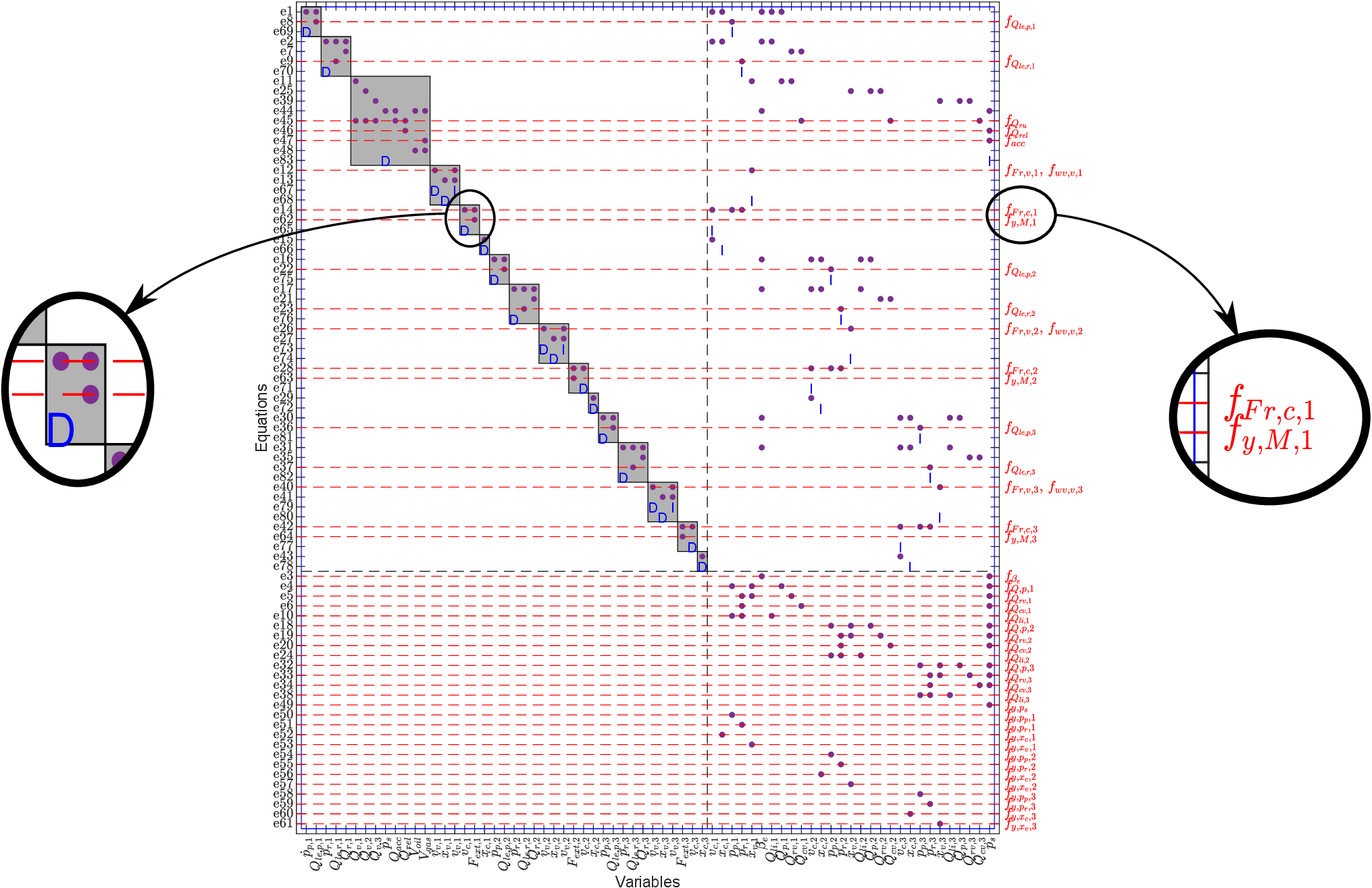}
\caption{DM Decomposition for the full plant, measurement of $F_{ext}$.\label{dm_full_fext}}
\end{figure}

The faults related to friction in the three cylinders are the only faults that appear in the just-determined part. Thus, they are not structurally detectable. Such a result could be expected, as each term appears in the same equations where the unknown variable $F_{ext}$ is present. When the friction fault is active, there is insufficient information to distinguish between the fault and a variation of the disturbance term. Furthermore, trying to estimate the external force acting on the cylinder is not feasible from a practical point of view, as reliable sensors aiming to measure it on an actual turbine are characterized by excessive cost. Furthermore, assuming it is possible to directly measure it, with a sensor that could get faulty, the DM decomposition in Figure~\ref{dm_full_fext} shows that the friction faults become detectable but not structurally isolable from the new faults introduced by the additional sensors.

All the other faults are contained in constraints that belong to the over-determined part of the system. Therefore, they are detectable. Furthermore, most of the faults are located in different equivalent classes, making them isolable. The two groups of faults for which this is not true are the valve faults and the supply line faults.

Each couple of faults affecting the PV appear in the same equation (see Equation~\ref{eq_xv}), making it impossible to structurally isolate them. As mentioned in Section~\ref{ch2_2}, the two faults have been inserted separately to highlight their different nature. The two faults are group-wise isolable with respect to other faults, meaning that is structurally possible to isolate the valve as a faulty component. Additionally, the model focuses on the description of the hydraulic system and doesn't account for the electrical dynamics of the valve's actuator, which is significantly faster and is considered as unmodeled dynamics in an analysis of structural relations. Finally, the faults in $Q_{p}$ and $Q_{rv}$ may also be affected by $f_{wv,v}$. This piece of information can be leveraged during the diagnostic system design.

The other faults that are group-wise isolable are related to the supply line, i.e., $f_{acc}$,$f_{Q_{ru}}$, and $f_{Q_{rel}}$. The oil leakages in the rotary unit and in the relief valve are acting on the same part of the system, i.e., the supply line. Thus, they are expected not to be distinguishable. The gas leakage in the accumulator is affecting Equation~\ref{eq_qacc}, through~\ref{eq_ps} and~\ref{eq_vgas}, on which both the other two faults are acting. The accumulator fault is a slow-varying fault. In practice, it can be detected and isolated by estimating the gas pre-charge pressure \citep{linigerSignalBasedGasLeakage2017}. As a consequence, that fault could be removed from the analysis and analyzed in this different framework. The other two faults can be isolated in practice by placing two pressure sensors, for example, before and after the rotary unit.

\hl{The simplification in the effective bulk modulus affects the structure of the system, by lowering the number of unknown parameters in the model. The choice improves the detectability/isolability properties of the system, thanks to the higher number of redundant relation that are consequent. As anticipated in the previous section, the price to pay is the impossibility of distinguishing the exact location where the oil is degrading.}

\hl{Since no substantial unmodelled dynamics have been reported in the well-established literature for the hydraulic actuator system, the dominant source of uncertainty in the model is associated with variations in the parameters that appear in its equations. Although parametric uncertainty may affect the performance of the residuals that can be derived based on the result of the structural analysis, the analysis itself, and hence, the diagnostic capabilities of the system, are invariant to such model mismatch. This is due to the fact that such analysis relies on the \emph{structural} properties of the constraints. The robustness of the presented algorithms with respect to parametric uncertainty is discussed later in Section \mbox{\ref{new5}}.}

\subsection{Distinct Operation Regions}\label{ch3_2}

The hydraulic pitch system's dynamics is characterized by different operating regions, because of the presence of hydraulic valves and the accumulator. The flow of each valve depends on the position of the related spool. The check valve, the relief valve, and the volume of gas in the accumulator depend on the relation between the system's pressure and different pressure values according to their characteristics. A table resuming the equations of interest and the inequality conditions is stated in Table~\ref{tab_cond}.

\begin{table}[H] 
\caption{Summary for distinct operation regions.\label{tab_cond}}
\centering
\small
\begin{tabular}{cccc}
\toprule
\textbf{Operation Regions} & \textbf{Equations}	& \textbf{Variables} & \textbf{Condition}\\
\midrule
1,2 & \ref{eq_qp}, \ref{eq_qrv}		& $x_v$ & $x_v\lessgtr0$\\
3,4 & \ref{eq_qcv}		& $p_r$, $p_s$ & $p_r\lessgtr p_s - p_{cv,c}$\\
5,6 & \ref{eq_qrel}		& $p_s$ & $p_s\lessgtr p_{cr,r}$\\
7,8 & \ref{eq_vgas}		& $p_s$ & $p_s\lessgtr p_{gas,0}$\\
\bottomrule
\end{tabular}
\end{table}

The different cases for each operating condition have been condensed into a single equation by leveraging Heaviside functions. This way, structural analysis was performed on the whole plant, and the result delivered the fault detectability/isolability capabilities throughout all the operating regions.

In this section, each of them is analyzed to determine whether any of the regions is limiting the performance of the system as a whole. The analysis is carried out by considering a simplified setup in which only one valve-cylinder pair is analyzed. In this way, the number of possible combinations is drastically reduced from $4^4=256$ to $2^4=16$. In fact, each of the additional subsystems is unrelated to each other, i.e., none of their variables can be used to match any variables with another actuation subsystem.

Firstly, the DM decomposition is applied to the simplified system as a whole. The result represents the benchmark for comparing the outcome of the different cases. For a quick visual comparison between the different regions, the toolbox provides an additional plot defined as 'isolability matrix'. If the (i,j) element is not null, then fault j is diagnosed when fault i occurs. As a consequence, the values only present on the diagonal are fully isolable, and the ones grouped in squares around the diagonal element are groupwise isolable.

\begin{figure}[t]
\centering
\includegraphics[width=0.9\textwidth]{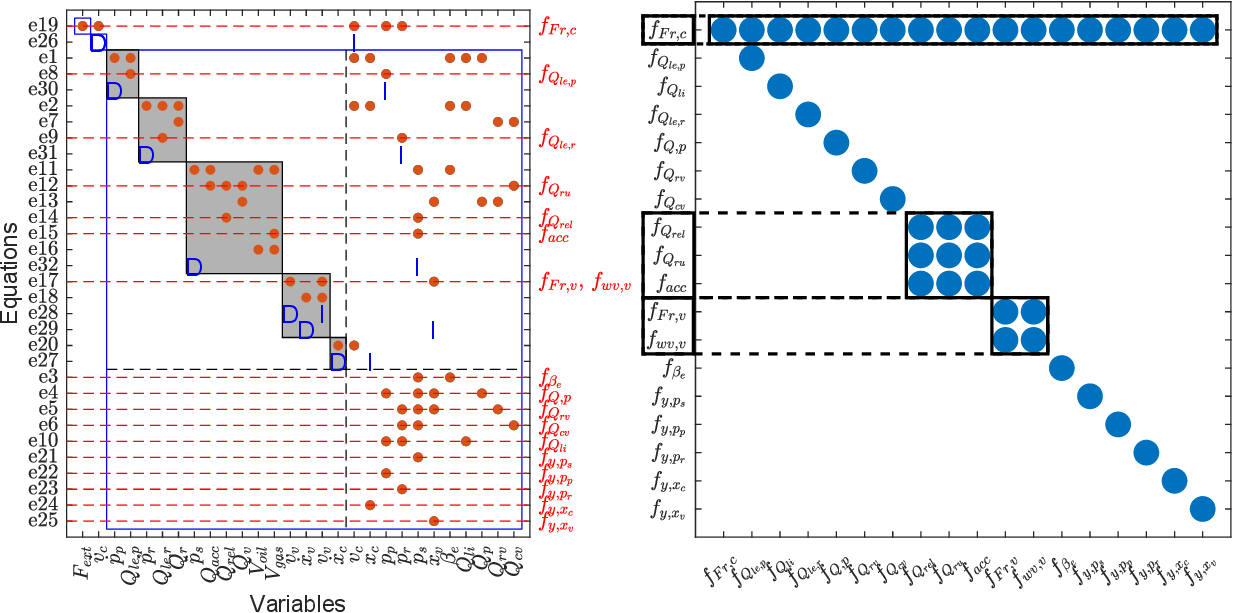}
\caption{Structural Analysis for plant with single cylinder: (\textbf{a}) DM Decomposition. (\textbf{b}) Isolability Matrix: $f_{Fr,c}$ is not detectable, faults $f_{Q_{rel}}, f_{Q_{ru}}, f_{acc}$ and $f_{F_{r,u}}, f_{wv,v}$ are group-wise isolable and all the other faults are isolable.\label{dm_iso_single}}
\end{figure} 

Non-isolable faults have non-zero elements on the related column. Non-detectable faults are instead placed row-wise to distinguish them from the non-isolable ones. Both the DM result and the related isolability matrix are stated in Figure~\ref{dm_iso_single}. As expected, the result follows the one in Section~\ref{ch3_1}. The two faults acting on the valve cannot be isolated, and the same applies to the trio $f_{acc}$,$f_{Q_{ru}}$, and $f_{Q_{rel}}$. The cylinder friction fault still can't be detected. The same results are highlighted on the isolability matrix.

\begin{figure}[h]
\centering
\includegraphics[width=0.88\textwidth]{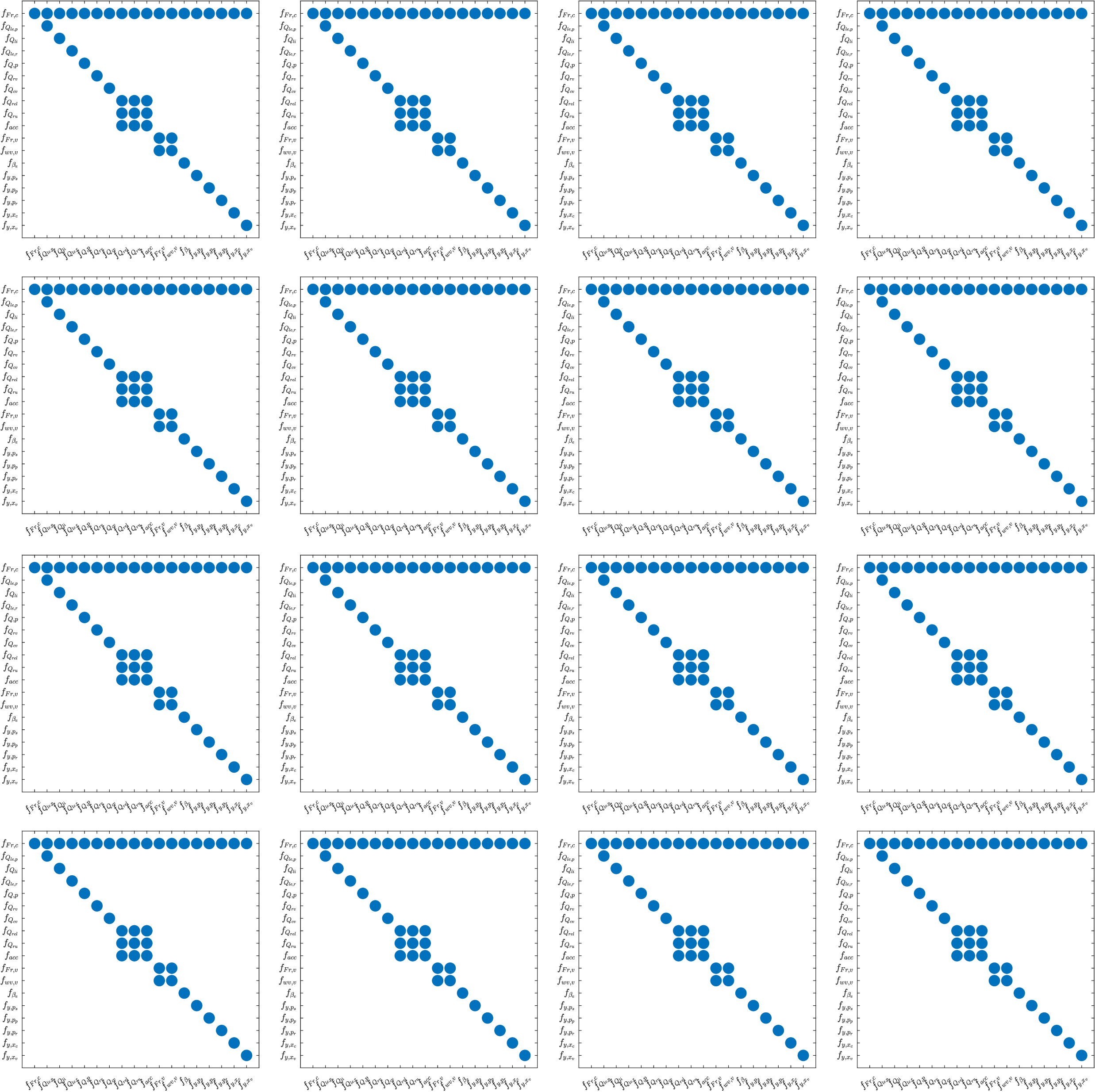}
\caption{Isolability Matrices for all Operation Regions. Detectability and isolability properties are independent of the operating regions. \label{iso_reg}}
\end{figure}

The analysis of the separate operation regions follows the same procedure as the previous one, where the equations containing a Heaviside function are replaced with the one characterizing each specific case. The results are presented in the form of fault isolation matrices, as they represent the most straightforward tool for comparison, in  Figure~\ref{iso_reg}. The pattern of every plot coincides with every other and with the one in Figure~\ref{dm_iso_single}. The outcome is that the analysis performed in Section~\ref{ch2} is meaningful for all the different conditions the plant is operating in. The full model can be utilized for deriving conclusions about sensor placement for improved diagnostic performance. 

%%%%

\section{Sensor placement for topology improvement}\label{ch4}

The results presented in Section~\ref{ch2} constitute a benchmark for the detectability and isolability properties of the system. Sensors have been placed such that they measure all the variables that are measured in practice. Each set of sensors measuring a different variable is defined as a sensor set, which is present for each of the turbine blades.

Generally, by increasing a system's redundancy thanks to the addition of more sensors, it is possible to improve the detectability and isolability properties of it. However, this is not the case for the topology under analysis, where the impossibility of detecting or isolating certain faults originates from the system structure and cannot be overcome by duplicating existing sensors. Other quantities could, in principle, be measured. However, placing such sensors is not feasible in real systems from an economic standpoint. Thus, they are not included in the analysis. On the other hand, it is possible that the performance presented in Section~\ref{ch2} can be obtained with a lower degree of sensor redundancies.

In this section, different sensor combinations are investigated to determine whether it is possible to achieve the same standard model performance, by minimizing the number of sensors to be installed on the system. The tool that has been applied is still the Structural Analysis Toolbox \citep{friskToolboxAnalysisDesign2017}, by gradually adding sensors and evaluating the performance.

The plant without any sensors is translated into a model containing an under-determined part. No fault is neither detectable nor isolable, as expected. Each sensor set is added and only the full sets $\{p_{p_1},p_{p_2},p_{p_3}\}$, $\{p_{r_1},p_{r_2},p_{r_3}\}$ and $\{x_{c_1},x_{c_2},x_{c_3}\}$ make the system just-determined. Moreover, the combination of the two pressure measurements leads to the same result, as long as one measurement per cylinder is provided. For example $\{p_{p_1},p_{p_2},p_{r_3}\}$.

\begin{figure}[t]
\centering
\includegraphics[width=0.9\textwidth]{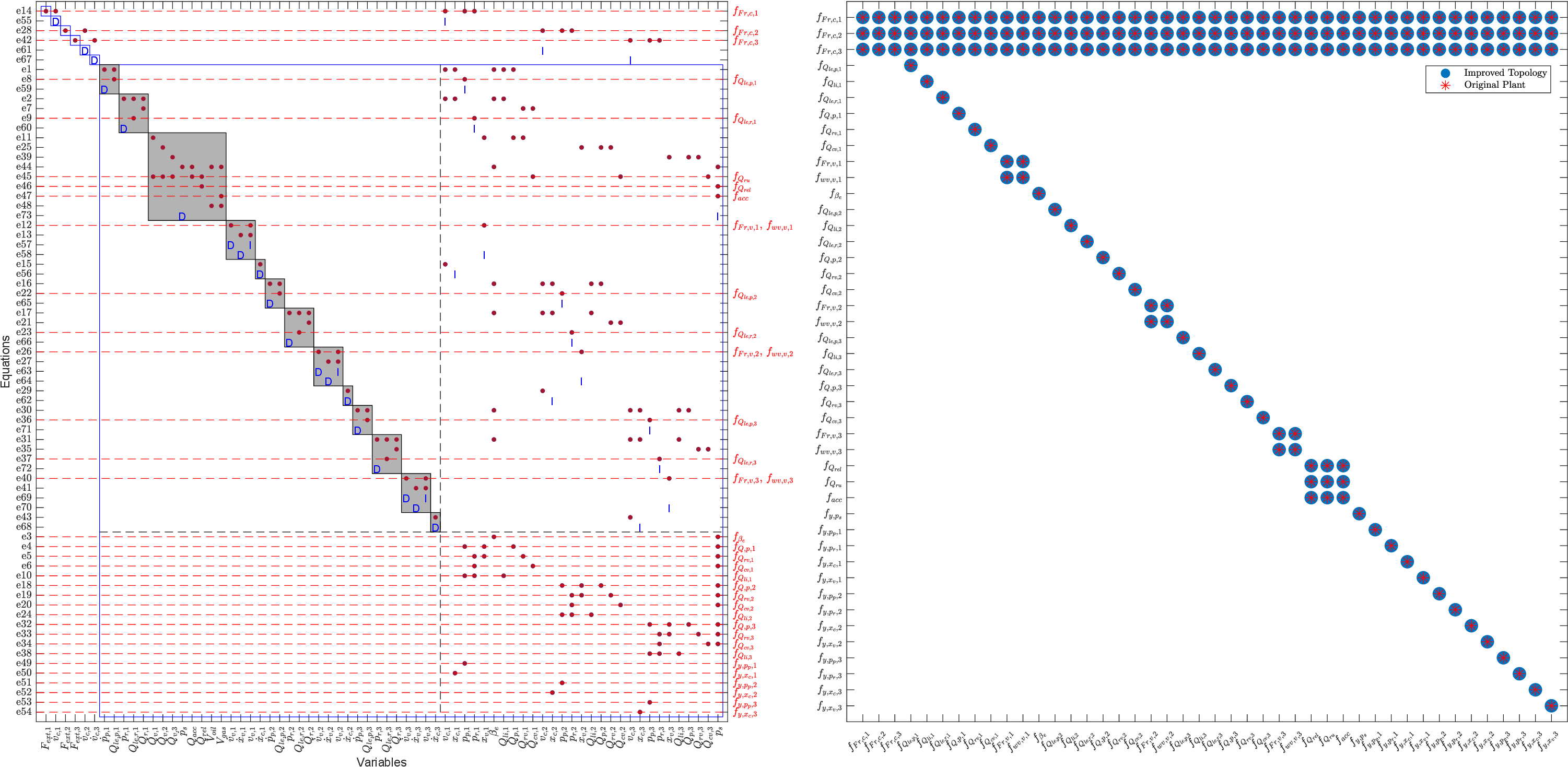}
\caption{Structural Analysis for plant with less sensors: (\textbf{a}) DM Decomposition. (\textbf{b}) Isolability Matrix.\label{dm_iso_red}}
\end{figure}

For each of the three just-determined models, each possible set of sensors is added. The combination of $\{x_{c_1},x_{c_2},x_{c_3}\}$ with either $\{p_{p_1},p_{p_2},p_{p_3}\}$ or $\{p_{r_1},p_{r_2},p_{r_3}\}$ delivers isolability performances equivalent to the ones of the benchmark model.

The combination of each of $\{p_{p_1},p_{p_2},p_{p_3}\}$ and $\{p_{r_1},p_{r_2},p_{r_3}\}$ with each other, or with any other set between $p_s$ and $\{x_{v_1},x_{v_2},x_{v_3}\}$, does not lead to any performance improvement. The same applies for the combination of $\{x_{c_1},x_{c_2},x_{c_3}\}$ with $p_s$ and/or $\{x_{v_1},x_{v_2},x_{v_3}\}$.

The outcome of the analysis is that the maximum structural isolability performance for such a plant is obtained by combining the measurement of the position of each of the cylinders ($\{x_{c_1},x_{c_2},x_{c_3}\}$), with the pressure in one of the two chambers, for each piston ($\{p_{p_1},p_{p_2},p_{p_3}\}$, $\{p_{r_1},p_{r_2},p_{r_3}\}$, or a correct combination of them). All the possible minimal sensor sets are summarized in Table~\ref{tab_sens}.

The DM decomposition and the isolability matrix for both the original and improved topology are stated for completion in Figure~\ref{dm_iso_red}. In this particular case, cylinder position and piston-chamber pressures are being measured.

\begin{table}[h] 
\caption{Summary of minimal sensors set.\label{tab_sens}}
% \newcolumntype{C}{>{\centering\arraybackslash}X}
\begin{tabular}{cp{7cm}}
\toprule
\textbf{Sensor Sets} & \textbf{Description}\\
\midrule
$\{x_{c_1},x_{c_2},x_{c_3}\}$, $\{p_{p_1},p_{p_2},p_{p_3}\}$& Cylinder pistons positions, all piston-side pressures\\
$\{x_{c_1},x_{c_2},x_{c_3}\}$, $\{p_{p_1},p_{p_2}\}$, $\{p_{r_3}\}$& Cylinder pistons positions, piston-side and rod-side pressures \\
$\{x_{c_1},x_{c_2},x_{c_3}\}$, $\{p_{p_1}\}$, $\{p_{r_2},p_{r_3}\}$&Cylinder pistons positions, piston-side and rod-side pressures\\
$\{x_{c_1},x_{c_2},x_{c_3}\}$, $\{p_{p_1},p_{p_3}\}$, $\{p_{r_2}\}$& Cylinder pistons positions, piston-side and rod-side pressures \\
$\{x_{c_1},x_{c_2},x_{c_3}\}$, $\{p_{p_2}\}$, $\{p_{r_1},p_{r_3}\}$&Cylinder pistons positions, piston-side and rod-side pressures\\
$\{x_{c_1},x_{c_2},x_{c_3}\}$, $\{p_{p_2},p_{p_3}\}$, $\{p_{r_1}\}$& Cylinder pistons positions, piston-side and rod-side pressures \\
$\{x_{c_1},x_{c_2},x_{c_3}\}$, $\{p_{p_3}\}$, $\{p_{r_1},p_{r_2}\}$&Cylinder pistons positions, piston-side and rod-side pressures\\
$\{x_{c_1},x_{c_2},x_{c_3}\}$, $\{p_{r_1},p_{r_2},p_{r_3}\}$& Cylinder pistons positions, all rod-side pressures\\
\bottomrule
\end{tabular}
\end{table}

%%%%

\section{Robustness to model uncertainty} \label{new5}

\hl{The main source of model uncertainty is related to the model parameters. Parameter values and uncertainty ranges are shown in \mbox{Table~\ref{tab:par_uncertainty}}. Additionally, the function $K(x_v)$ in Equations \mbox{\ref{eq_qp}} and \mbox{\ref{eq_qrv}} has an uncertainty of $5\%$.}

\begin{table}[h] 
    \centering
    \small
    \begin{tabular}{c|p{5cm}|c|c|c}
       Symbol  & Description & Value & Units & $\pm\,\%$ \\
         \hline
         \TT
       $M_{eq}$  & Cylinder's piston equivalent mass & 104 & $Kg$ & 1  \\
       $A_p$  & Piston side cylinder's cross section & 0.015 & $m^2$ & 1 \\
       $A_r$  & Rod side cylinder's cross section & 0.009 & $m^2$ & 1 \\
       $B_v$ & Viscous friction coefficient & 6500 & $Nms^{-1}$ & 80 \\
       $F_c$ & Coulomb friction coefficient & 2600 & $N$ & 80 \\
       $\gamma$ & Sign function approximation & $2 \times 10^{-4}$ & $-$ & 0 \\
       $\beta_{e,0}$ & Oil's effective bulk modulus & $7 \times 10^{8}$ & $Pa$ & 40 \\ 
       $V_{0,p}$  & Piston side dead volume & $3.7 \times 10^{-4}$ & $m^3$ & 1 \\
       $V_{0,r}$  & Rod side dead volume & $1.5 \times 10^{-4}$ & $m^3$ & 1  \\
       $x_{c,max}$  & Cylinder's stroke & 1.35 & $m$ & 1 \\
       $C_{le,p}$  & External leakage coefficient (piston) & 0 & $m^3sKg^{-1}$ & 0 \\
       $C_{le,r}$  & External leakage coefficient (rod) & 0 & $m^3sKg^{-1}$ & 0 \\
       $C_{li}$  & Internal leakage coefficient & 0 & $m^3sKg^{-1}$ & 0 \\
       $\phi_v$ & Cross sections ratio & 0.5867 & $-$ & 5 \\
       $\xi$ & Valve's Damping ratio & 1 & $-$ & 1 \\
       $\omega_0$ & Valve's Natural frequency & 440 & $rad s^{-1}$ & 1 \\
       $K_{cv}$  & Check valve flow coefficient & $1.5 \times 10^{-8}$ & $m^3sKg^{-1}$ & 10  \\
       $p_{cv,c}$  & Check valve crack pressure & $0.5 \times 10^{5}$ & $Pa$ & 50  \\
       $k$  & Accumulator's gas stiffness & 1.4 & $m^3sKg^{-1}$ & 0 \\
       $K_{rel}$  & Relief valve flow coefficient & $1.5 \times 10^{-8}$ & $m^3sKg^{-1}$ & 10  \\
       $p_{cr,r}$  & Relief pressure & $2.1 \times 10^{7}$ & $Pa$ & 0.5 \\
       $V_{acc}$  & Accumulator volume & 0.1 & $m^3$ & 1 \\
       $V_{hose}$  & Hose volume & $1 \times 10^{-3}$ & $m^3$ & 1 \\
       $p_{gas,0}$ & Accumulator nominal pressure & $1.89 \times 10^{7}$ & $Pa$ & 0.5 
       \BB\\       
         \hline
    \end{tabular}
    \caption{\hl{Table of parameters with their values and uncertainty ranges (in percentage).}}
    \label{tab:par_uncertainty}
\end{table}

\subsection{\hl{Model uncertainty}}
\hl{Denote the modelled behaviour by $\hat{g}$, and the actual by $\breve{g}$, then modelled and actual residuals become parameterised in $g$ as,}
\begin{align}
    \hat{r}(t) &= \hat{g}(\{\hat{c}(\cdot), u(t), y(t)\} \in {\mathcal{M}}_1) 
    \label{residual ideal} \\
    \breve{r}(t) &= \breve{g}(\{{c}(\cdot), d(t), f(t), u(t), y(t)\} \in {\mathcal{M}}_1) 
    \label{residual real}
\end{align}
\hl{where $d$ represent disturbances, $u$ is input, $y$ the measurements and $f$ the signals representing faults.
It should be noted that in the absence of faults and disturbances the actual and model residual signals are equal, i.e. $\breve{r}(t) = \hat{r}(t)$ when $d(t) = 0, \; f(t) = 0$.}

\hl{Perturbation analysis then provides the sensitivity of the residual to uncertainty of parameters, i.e. to a variation of the parameter vector $\Delta p = \breve{p}-\hat{p}$. Indeed, using Taylor expansion around the values used in the model for the parameters, disturbances and the considered fault, the actual residual $\breve{r}$ is approximated as:}
\begin{align} \label{sensitivity to parameter change}
 \breve{r}(t) &\approx \underbrace{\frac{\partial \breve{r}}{\partial y} \Delta y + \frac{\partial \breve{r}}{\partial u} \Delta u}_{\hat{r}(t)} + \underbrace{\frac{\partial \breve{r}}{\partial p} \Delta p + \frac{\partial \breve{r}}{\partial d} d + \frac{\partial \breve{r}}{\partial f}  f}_{\tilde{r}(t) \triangleq \breve{r}(t) - \hat{r}(t)}
\end{align}
\hl{and the consequences of model uncertainty may include:}
\begin{itemize}
    \item \hl{the residual becomes sensitive to disturbances, whereas, by design of $r$, the signals $d$, $u$ and $y$ do not affect the residual in the nominal case.}
    \item \hl{some faults $f$ may be masked by $\Delta p$ and $d$.}
\end{itemize}
\hl{The latter implies that when evaluating $\hat{r}(t)$ by comparing single samples with a threshold $T$ in the noise free case, safe conclusions on the detection of a fault $f$ can be drawn if the following condition holds}
\begin{equation} \label{eq:condition_bounds}
    \Vert f \Vert_{\infty} > \frac{\left \Vert \displaystyle \frac{\partial \breve{r}}{\partial p} \right \Vert_{\infty} \Vert \Delta p \Vert_2 + \left \Vert \displaystyle \frac{\partial \breve{r}}{\partial d} \right \Vert_{\infty} \Vert d \Vert_{\infty}}{\left \Vert \displaystyle \frac{\partial \breve{r}}{\partial f} \right \Vert_{\infty}}
\end{equation}
\hl{where the $\mathcal{L}_{\infty}$ norm of a time-varying vector (equivalently matrix) $y(t)$ is defined as $\Vert y \Vert_{\infty} \triangleq \sup\limits_t \Vert y(t) \Vert$ using any vector (equivalently induced matrix) norm.}

\subsection{\hl{Example}}
\hl{As an example, a residual associated with the results from the structural analysis in Section 3 is given in analytical form by}
\begin{align}
    \breve{r} = &-\frac{\dot{p}_r}{B_e}[V_{or}-A_r(x_c-x_{c,max})] + A_r \dot{x}_c \blt{-f_{Qle,r}} \nonumber\\
    &+ \phi_v  K_v(x_v)  \sqrt{|p_r - p_s|}  \textit{sign}(p_r - p_s)  \textit{H}(-x_v) \nonumber\\
    &+ K_{cv}(p_s- p_r -p_{cv_c} )  \textit{H}(p_s-p_r-p_{cv_c} )  
\end{align}

\hl{Consider an external leakage fault $f = f_{Qle,r}$. The bulk modulus $B_e$ and the check valve flow coefficient $K_{cv}$ are uncertain parameters with variations $\vert \Delta B_e \vert < \delta_1$ and $\vert \Delta K_{cv} \vert < \delta_2$. The bounds $\delta_1, \delta_2 > 0$ are known from \mbox{Table~\ref{tab:par_uncertainty}}. Following the notation introduced in \mbox{\eqref{eq:condition_bounds}} with $p \triangleq \begin{bmatrix}
    B_e & K_{cv}
\end{bmatrix}^T$ and $\Delta p \triangleq \begin{bmatrix}
    \Delta B_e & \Delta K_{cv}
\end{bmatrix}^T$ one gets}
\begin{align}
    \frac{\partial \breve{r}}{\partial p} &= \begin{bmatrix}\nonumber
        \displaystyle \frac{\dot{p}_r}{B_e^2}[V_{or}-A_r(x_c-x_{c,max})] & (p_s- p_r -p_{cv_c} )  \textit{H}(p_s-p_r-p_{cv_c} )
    \end{bmatrix}\\
    \frac{\partial \breve{r}}{\partial d} &= 0\\
    \frac{\partial \breve{r}}{\partial f_{Qle,r}} &= -1\;.\nonumber
\end{align}

\hl{Let $\phi > 0$ be a bound on $\displaystyle \left \Vert \frac{\partial \breve{r}}{\partial p} \right \Vert_{\infty}$, i.e.}
\begin{equation}
    \sup\limits_t \sqrt{\displaystyle \frac{\dot{p}^2_r}{B_e^4}[V_{or}-A_r(x_c-x_{c,max})]^2 + (p_s- p_r -p_{cv_c} )^2} \leq \phi.
\end{equation}

\hl{Then application of condition \mbox{\eqref{eq:condition_bounds}} leads to the requirement}
\begin{equation}
    \Vert f_{Qle,r} \Vert_{\infty} > \phi \sqrt{\delta_1^2 + \delta_2^2}\; .
\end{equation}

\hl{The presence of a fault is detected if $\vert \hat{r}(t) \vert > \phi $, which ensures that the detection of the fault is robust to parameter variations. Estimating $\phi$ is subject to the values obtained by the states and outputs of the system, i.e., to excitation conditions and operating mode. This suggests that such bounds can often be conservative even if $\delta_1, \delta_2$ are not.}

\section{Discussion}\label{ch5}

Structural analysis was used to obtain an overview of the fault diagnosis capabilities that can be achieved with the standard sensor configuration of hydraulic pitch control systems. An initial analysis was carried out by including all the possible sensors that such systems may include, i.e., pressure sensors for the supply circuit and both chambers of each cylinder, position sensors for each cylinder, and position sensors for each valve.

In terms of isolability, two groups of violations of constraints were shown to be group-wise isolable. The first group was represented by changes in voltage or friction in the valve (Equation~\ref{eq_xv}). The second referred to leakages on the supply line, i.e., rotary unit leakage, relief valve defect, or accumulator defect. All other defects that were inserted into the system were shown to be isolable, which means that standard residual generators will suffice to detect and isolate the root cause of the changes they create in the pitch system signals.

The outcome of the detectability analysis was that the only fault that couldn't be detected from a structural perspective is an increased friction in the cylinder. This was shown to be due to the presence of a disturbance term in the same equation. This result from the structural analysis can be used to automatically generate residuals that achieve very convincing coverage of detectability and isolability, and it points to the one component in the system where supplemental effort is needed for condition monitoring. The structural result does not imply that changes in cylinder friction could not be detected. It merely shows that the residuals generated by structural analysis will not discriminate changes in the cylinder friction from variations in load torques exerted on the turbine blades from wind. The estimation of cylinder friction is essential for pitch system condition monitoring and has been studied in \citep{dallabonaFrictionEstimationCondition2024a}.
Finally, it should be noted that the topology of a system may change in different operational modes, or according to system excitation via different paths of control. This can be utilized to enhance fault isolation over what is visible from the standard structural isolability analysis.

Another approach to improve on the effectiveness of the structure-based diagnosis, pertains to \emph{active fault isolation}, i.e., a set of techniques where control inputs are excited and/or operational mode is changed, to explore different input-to-output paths in topology. Structural results for active fault isolation were presented in \citep{blankeStructuralDesignSystems2007}, and enhanced isolation by shifting between operational modes, was shown for a water-hydraulic medical "water for injection" plant in \citep{laursenFaultDiagnosisWater2008a}. Methods for signal-based active isolation enhance the structural results by using the time/frequency domain properties, which structural analysis does not necessarily show. A generic setup for active fault diagnosis was published in \citep{poulsenActiveFaultDiagnosis2008}, and tested on a high fidelity wind turbine simulation in \citep{niemannFaultDiagnosisCondition2018} where continuous-time analysis and linear system assumptions were utilized.

\hl{As discussed in \mbox{Section~\ref{new5}}, the robustness of fault detection can be compromised by parametric and model uncertainty. Condition \mbox{\eqref{eq:condition_bounds}} relies on the calculation of bounds on conservative bounds on the gradients $\frac{\partial \breve{r}}{\partial p}$, $\frac{\partial \breve{r}}{\partial d}$ and $\frac{\partial \breve{r}}{\partial f}$, which could limit what can be detected. Methods based on signal processing and statistical change detection offer supplementary features that enhance diagnosability of faults even in the presence of noise. For example, a test statistics $g_r$ of $r(t)$ can be compared with a threshold $h$ that is obtained by imposing requirements on probabilities of detection and false alarms \mbox{\cite{blankeDiagnosisFaultTolerantControl2016}}. Additional robustness can be achieved without sacrificing sensitivity by using commissioning data to obtain adaptive thresholds.}

%%%%

\section{Conclusions}\label{ch6}

This article provided an extended model that can be used for fault diagnosis and fault tolerant control in hydraulic pitch systems of wind turbines. A wide range of major faults that can occur in the system's lifetime, were integrated into the model that corresponds to dynamics during nomimal operation. This integration comprised information on the faults' signal properties as well as their change profiles. The combination of these two pieces of information was leveraged for deriving the structural detectability and isolability properties that are achievable with the presented configuration, i.e., by means of a Dulmage-Mendelson decomposition. Different sensor configurations were tested to reduce the level of redundancy while maintaining the same isolability performances. The outcome was an indication of the minimal sensor requirement for facilitating maximum fault structural isolability in the most generic scenario of wind turbine hydraulic pitching. \hl{Finally, elements of the robustness of the presented approach with respect to parametric uncertainty were discussed in a test cases. The results highlighted that although the quality of fault detection and estimation is subject to variations in the model parameters, the latter do not affect the structural detectability and isolability features of the system.}

\hl{Future work will include a full-scale sensitivity analysis to quantify the effects of parametric uncertainty on the diagnostic performance of the generated residuals. Furthermore, active fault diagnosis will be considered in connection to potential improvements on the detection and isolation capabilities of the system. Experimental verification of the results will be pursued by implementing structure-based residual generators on a real pitch hydraulic system.}

%%%%

\section*{CRediT autorship contribution statement}
This paper is a collaboration between authors. A.D. is the main contributor and is responsible for the conceptualization, formal analysis, investigation, and most of the writing of the paper. M.B., H.C.P. and D.P. contributed to the writing and with in-depth reviews throughout the process.  All authors have read and agreed to the published version of the manuscript.

\section*{Declaration of competing interest} The authors declare no conflict of interest.

\section*{Acknowledgements}
This research was funded by The Danish Energy Technology Development and Demonstration Programm (EUDP) through the project: “Decreased Cost of Energy (CoE) from wind turbines by reducing pitch system faults”, grant number 64022-1058. The authors much appreciate this support.





\bibliographystyle{abbrv}
\bibliography{2024-pitchdiag-bibl}




% \end{thebibliography}
\end{document}